\documentclass[12pt]{article}           
       
\def\title{Why is the Legendre Transformation Involutive?}

\long\def\abstract{ 
The question posed in the title is answered in terms of a simple
pictorial argument that is manifestly symmetric between the two 
functions 
that are Legendre transform of each other.
}

\input epsf

\usepackage{amssymb,latexsym}              \catcode`\"=\active \let"=\"

\def\ifundefined#1{\expandafter\ifx\csname#1\endcsname\relax}
\def\bye{\end{document}}   
\long\def\new#1\endnew{{\bf #1}}
\long\def\del#1\enddel{} 

\def\HS#1 {\hspace*{#1pt}} \def\VS#1 {\vspace*{#1pt}}
\def\BC{\begin{center}}    
\def\EC{\end{center}}

\let\0=\over      \let\1=\vec      \def\2{{1\over2}}    \let\3=\ss
\let\4=\underline \let\5=\overline \let\6=\partial      \def\7#1{{#1}\llap{/}}
\def\8#1{{\textstyle{#1}}}         \def\9#1{{\ifmmode{\pmb{#1}}\else\bf#1\fi}}

          \def\({\left(}       \def\){\right)}

\def\eeql#1 {\label{#1}\eeq}        
\def\beq{\begin{equation}}      \def\eeq{\end{equation}}        
\def\bea{\begin{eqnarray}}      \def\eea{\end{eqnarray}}

\let\and=\wedge

\let\bra=\langle        \let\ket=\rangle        \def\<#1\>{\bra #1 \ket}

\def\rel#1 #2{\buildrel #1 \over {#2}}

\def\plb#1 #2 {Phys. Lett. {\bf B#1} #2 }
\def\phr#1 #2 {Phys. Rep. {\bf  #1} #2 }        
\def\npb#1 #2 {Nucl. Phys. {\bf B#1} #2 }
\def\aph#1 #2 {Ann. Phys. {\bf #1} #2 }         
\def\jmp#1 #2 {J. Math. Phys. {\bf #1} #2 }
\def\jgp#1 #2 {J. Geom. Phys. {\bf #1} #2 }
\def\prd#1 #2 {Phys. Rev. {\bf D#1} #2 }
\def\prl#1 #2 {Phys. Rev. Lett. {\bf #1} #2 }
\def\rmp#1 #2 {Rev. Mod. Phys.  {\bf #1} #2 }
\def\zpc#1 {Z. Phys. {\bf #1C} }
\def\cmp#1 #2 {Commun. Math. Phys. {\bf #1} #2 }
\def\cqg#1 #2 {Class.Quant.Grav. {\bf #1} #2 }
\def\mpl#1 {Mod. Phys. Lett. {\bf A#1} }
\def\cpc#1 {Computer Phys. Commun. {\bf #1} }   
\def\ijmp#1 {Int. J. Mod. Phys. {\bf A#1} }
\def\ijmpC#1 {Int. J. Mod. Phys. {\bf C#1} }
\def\atmp#1 {Adv. Theor. Math. Phys. {\bf #1} }

\def\BP{\begin{picture}} \def\EP{\end{picture}}         
\newcounter{TRefNX} \let\OLDcite=\cite  \makeatletter
\def\makeTRefs#1{\@for  \NewTRef:=#1\do{\global\makeTRef{\NewTRef}}}
\def\makeTRef#1{\ifundefined{TRef#1}\stepcounter{TRefNX}%
\expandafter\xdef\csname TRef#1\endcsname{\theTRefNX}\fi}\makeatother
\def\NEWcite#1{\makeTRefs{#1}\OLDcite{#1}}  

\ifundefined{draftmode} {} 
\else        
   \let\cite=\NEWcite
   \newcount\HOUR  \HOUR=\the\time \divide\HOUR by 60   \multiply \HOUR by 60 
   \newcount\MIN   \MIN=\the\time  \advance\MIN by -\HOUR  \divide\HOUR by 60
   \ifnum  \MIN>9  \def\printTIME{{\it\the\HOUR\,:\,\the\MIN}}
   \else   \def\printTIME{{\it\the\HOUR\,:\,0\the\MIN}} 
   \fi 
   
   \def\LLab#1{\BP(0,0)\unitlength=1mm\put(-12,.5){\makebox(0,0)[cr]{\small #1
        \rlap{$_{_{\makeatletter\csname TRef#1\endcsname\makeatother}}$}}}\EP}
\fi

\begin{document}


\begin{center} 
{\Large\bf   \title }\vskip 10mm
Harald Skarke\\[3mm]
Institut f\"ur Theoretische Physik, Technische Universit\"at Wien\\
Wiedner Hauptstra\ss e 8--10, A-1040 Wien, AUSTRIA\\
{\tt skarke@hep.itp.tuwien.ac.at}
        
\vskip 1 cm                  
{\bf ABSTRACT } 

\abstract

\rule{9cm}{.5mm}
\end{center}




  

The Legendre transformation is a mathematical concept of great 
significance to physics. In mechanics and field theory it provides the 
transition between Hamiltonian and Lagrangian descriptions, and in 
thermodynamics it relates the different potentials. 
\del
Nevertheless it is usually introduced just along the way\footnote{%
A notable exception is \cite{zia}.}, leaving the 
impression of a sleight of hand -- at least to the author of the present note, 
thereby providing the motivation for it.
\enddel
Nevertheless, with very few exceptions (notably \cite{zia}), it is usually 
introduced just along the way, leaving the 
impression of a sleight of hand.
The feeling that some essential point might be missing from the standard 
description provided the motivation for the present considerations.
In the following the definition of 
the Legendre transform $G(y)$ of a function
$F(x)$ and a simple argument for its involutivity will be given.

Let us assume that the function $F(x)$ is 
continuously differentiable, with a 
derivative
\[ f(x):= F'(x) \]
that is strictly monotonically increasing.
Then the function $f(x)$
can be inverted to $g(y)$:
\[ y = f(x) ~~~ \Longleftrightarrow  ~~~ x = g(y), \]
and the Legendre transform of $F(x)$ is defined as
\beq G(y) := [xy - F(x)]|_{x=g(y)} .\eeql{legtr}
If one performs the same operation on $G(y)$:
\[ z := G'(y),~~~H(z):= [yz - G(y)]|_{y=h(z)},\]
where $h$ is the function inverse to $G'$,
a very short calculation reveals
that $z=x$, $h=f$ and $H=F$,
i.e. one has returned to the original function.
This is, of course, perfectly sufficient as a proof of involutivity,
but a physicist would prefer a more intuitive explanation,
ideally in terms of geometry.
The standard geometric interpretation of the Legendre transform 
proceeds by considering the graph of the convex function $F(x)$ and its
tangents.
This is a correct pictorial account of formula (\ref{legtr}) which can be
used to give a geometric proof (see, e.g., Arnold \cite{arn}), 
but it does not make the 
symmetry between $F$, $f$ and $x$ on one side and $G$, $g$ and $y$ on the 
other side manifest.
Let us therefore look at the graph of the monotonic function $f(x)$ instead.

\begin{figure}[htb]
\epsfxsize=3in
\hfil\epsfbox{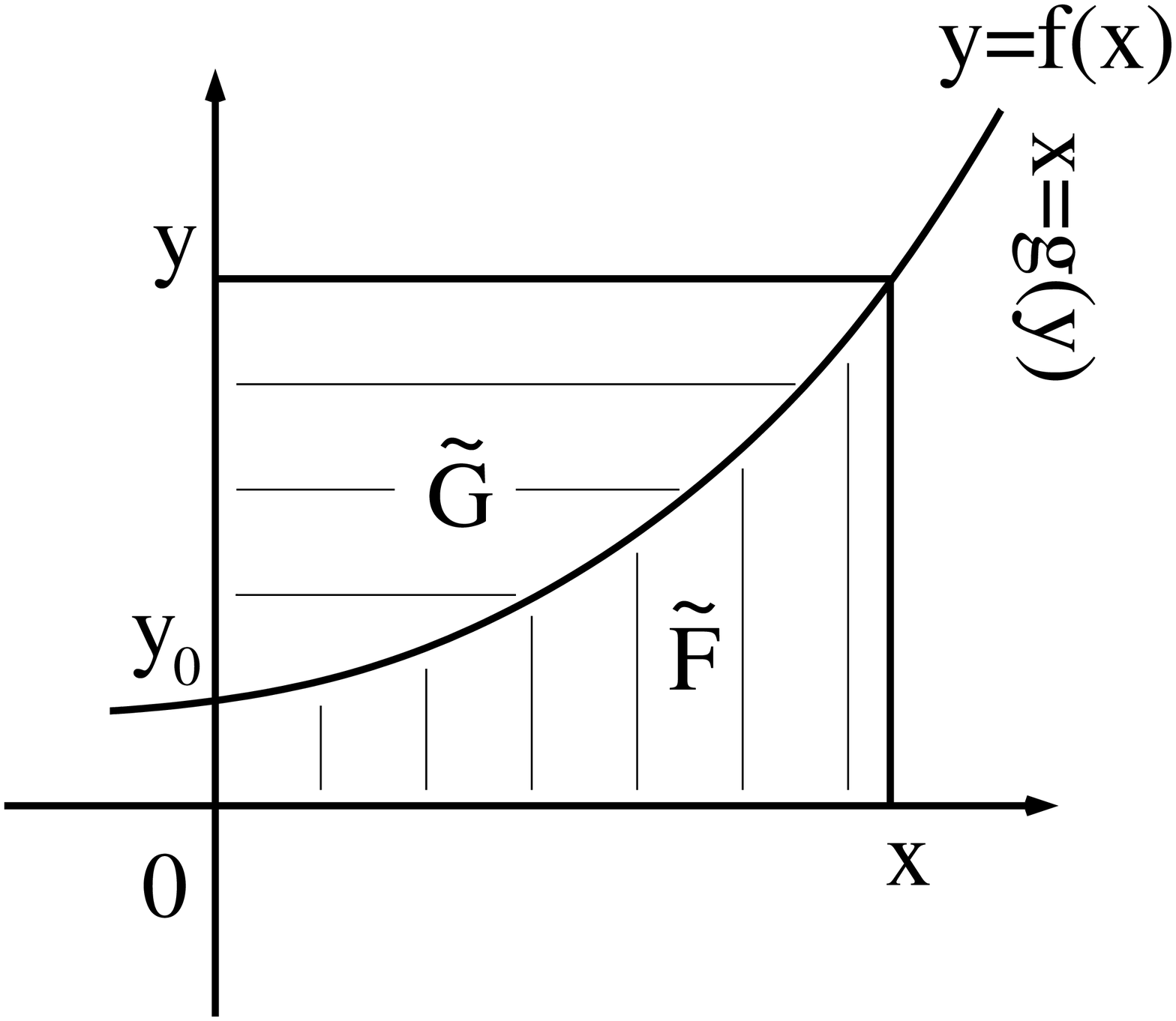}\hfil\epsfxsize=3in
\hfill\epsfbox{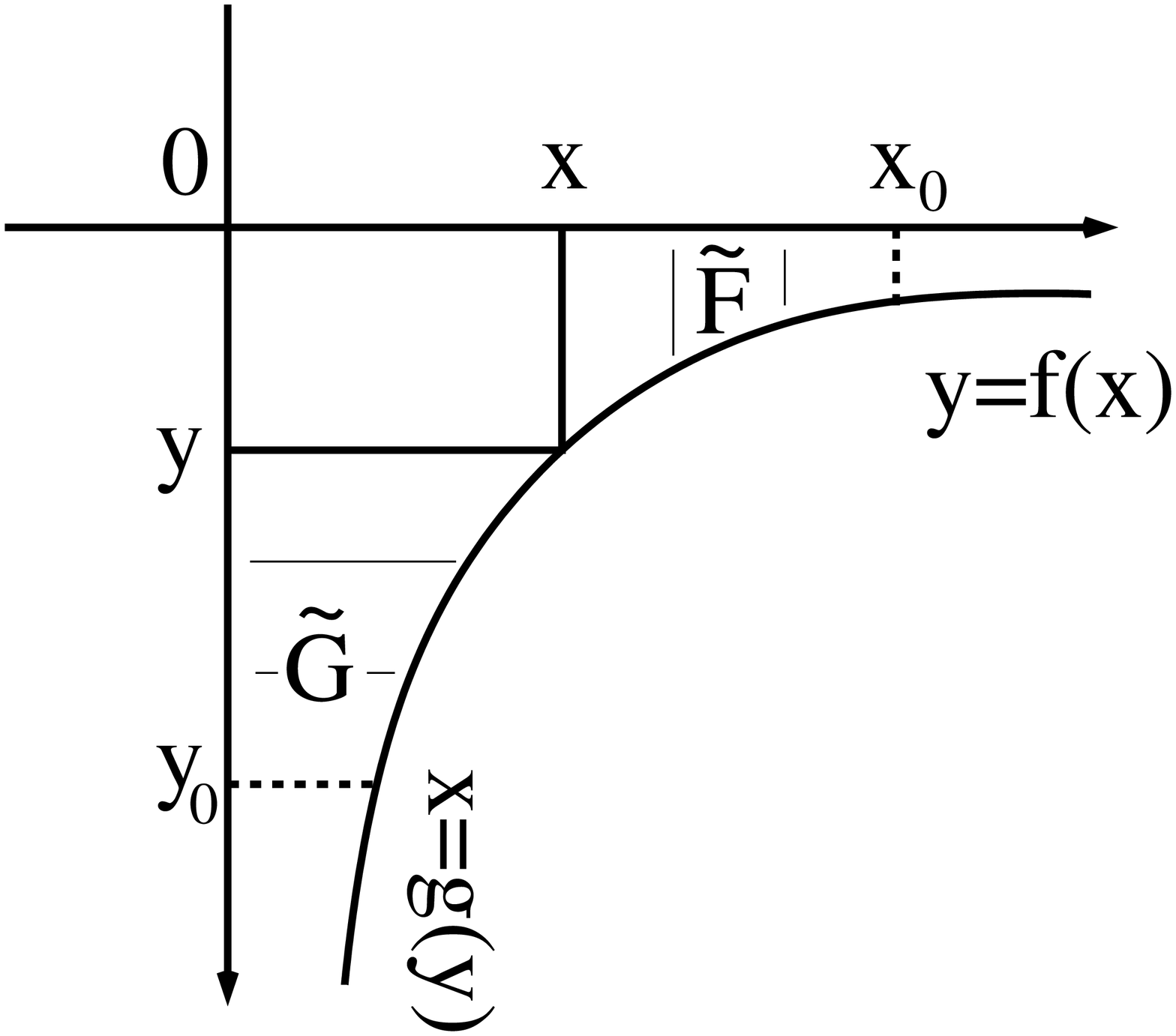}\hfil
\caption{The graph of $y=f(x)$ for the cases (a) $x>0, y>0$ and (b) $x>0, y<0$}
\label{fig:leg12}
\end{figure}

We first assume that $x$ and $f(x)$ are positive; see the first diagram
of Figure 1 which depicts the graph of $f(x)$ over $x$.
The same plot can be interpreted as the graph of $g(y)$ with respect to the 
$y$-axis.
Expressed in a symmetric manner, the figure shows the locus of all pairs 
$(x,y)$ with $y=f(x)$ or, equivalently, $x=g(y)$.
Now consider the rectangle 
bounded by the coordinate axes and their parallels through
such a point $(x,y)$.
The area of that rectangle is $A = xy$, and the graph cuts 
it into two parts with areas $\widetilde F$ and $\widetilde G$,
respectively.
From the figure it is immediately obvious that
\[
\widetilde F = \int_{x_0}^xf(\hat x)d\hat x,~~~~~
\widetilde G = \int_{y_0}^yg(\hat y)d\hat y,~~~~~
\widetilde F + \widetilde G = xy,
\]
with $x_0=0$ if the graph intersects the $y$-axis in $y_0\ge 0$, and $y_0=0$ 
if the graph intersects the $x$-axis in $x_0\ge 0$.
Clearly $\widetilde F$ is a function of $x$ with
$ \widetilde F' (x)  = f(x) = F'(x)$, hence 
\[ F(x) = \widetilde F(x) +c,~~~~~ G(y) = \widetilde G(y) -c   \]
for some real constant $c$.
So $F$ is, up to a constant, the area under the graph of $f$, and $G$ is,
up to minus that constant, the area under the graph of $g$, and the symmetry
is manifest.

What if our assumptions $x\ge 0$, $y\ge 0$ are not satisfied?
For $x\le 0$, $y\le 0$ the argument is essentially unmodified since 
$(-x)(-y)=xy$.
If $xy<0$ consider the second part of Figure 1.
Here we have fixed two arbitrary constant values $x_0$, $y_0$ in such a way
that $x_0>x>0$, $y_0 < y < 0$ for the range of pairs $(x,y)$ that we want to
consider.
Denote by $A_0$ the area determined by the coordinate axes, the vertical line 
through $x_0$, the horizontal line through $y_0$ and the graph.
Then we have 
\[ A_0 = -xy + \widetilde F + \widetilde G ~~~\hbox{with}~~~
\widetilde F = - \int_x^{x_0}f(\hat x)d\hat x,~~~
\widetilde G = \int_{y_0}^yg(\hat y)d\hat y.
\]
Up to the constant $A_0$ that can be absorbed in the redefinitions from 
$\widetilde F$ to $F$ and from $\widetilde G$ to $G$,
$\widetilde F$ and $\widetilde G$ again add up to $xy$.

\del
We note that the ambiguity in $F\sim \widetilde F$, $G\sim \widetilde G$, 
with `$\sim$' meaning
`equal up to a constant function', fits well with physical ambiguities
of how to define zeroes of potentials.
\enddel
The fact that the present picture requires redefinitions of functions
by constants is directly related to the interpretation of $F(x)$ and $G(y)$ as 
integrals of $f(x)$ and $g(y)$, respectively. 
As always, integrals are well-defined only up to equivalences of
the type $F\sim \widetilde F$, 
with `$\sim$' meaning `equal up to a constant function'.
This fits precisely with the physical interpretation, where the predictions 
do not change if quantities like the Hamiltonian or thermodynamic potentials 
are redefined by constants.

\noindent
{\bf Note added after completion:} I first presented this material in 
informal talks on March 15, 2012 in Vienna and on June 4, 2012 in Heidelberg.
After 
writing it up in the form 
of the present manuscript, I became aware of \cite{hoff}
which is dated June 29, 2012 (submission) / August 22, 2012 (publication)
and has some overlap in content.
I am grateful to Johanna Knapp for pointing out this reference to me.

\bye